\documentclass[
reprint,
superscriptaddress,
amsmath,
amssymb,
aps,
pra, 
longbibliography,
floatfix
]{revtex4-1}

\usepackage{graphicx}
\usepackage{xcolor}

\usepackage{tabularx}
\usepackage{dcolumn} 
\newcolumntype{d}[1]{D{.}{.}{#1}}
\usepackage{booktabs}

\usepackage{comment}
\usepackage{enumitem}

\begin{document}

\title{Layer and spontaneous polarizations in perovskite oxides and their interplay in multiferroic bismuth ferrite}

\author{Nicola A. Spaldin}
\affiliation{Department of Materials, ETH Zurich, CH-8093 Z\"{u}rich, Switzerland}
\author{Ipek Efe}
\affiliation{Department of Materials, ETH Zurich, CH-8093 Z\"{u}rich, Switzerland}
\author{Marta D. Rossell}
\affiliation{Electron Microscopy Center, Swiss Federal Laboratories for Materials Science and Technology, Empa, 8600, Dübendorf, Switzerland}
\author{Chiara Gattinoni}
\affiliation{Department of Materials, ETH Zurich, CH-8093 Z\"{u}rich, Switzerland}

\date {\today}

\begin{abstract}

We review the concept of surface charge, first in the context of the polarization in ferroelectric materials, and second in the context of layers of charged ions in ionic insulators. While the former is traditionally discussed in the ferroelectrics community, and the latter in the surface science community, we remind the reader that the two descriptions are conveniently unified within the modern theory of polarization. In both cases, the surface charge leads to electrostatic instability --- the so-called ``polar catastrophe'' --- if it is not compensated, and we review the range of phenomena that arise as a result of different compensation mechanisms. We illustrate these concepts using the example of the prototypical multiferroic bismuth ferrite, BiFeO$_3$, which is unusual in that its spontaneous ferroelectric polarization and its layer charges can be of the same magnitude. As a result, for certain combinations of polarization orientation and surface termination its surface charge is self-compensating. We use density functional calculations of BiFeO$_3$ slabs and superlattices, analysis of high-resolution transmission electron micrographs as well as examples from the literature to explore the consequences of this peculiarity. \\

\end{abstract}

\maketitle

\section{Introduction}

Bound charge at the surface of an insulator, or at an interface between two insulating materials, must be screened in order to avoid a so-called polar catastrophe caused by a divergence of the electrostatic energy~\cite{Stengel:2011}.
Such bound surface charge, $\sigma_{\text{surf}}$, exists whenever there is an uncompensated component of the bulk polarization, $\vec{P}_{\text{bulk}}$, perpendicular to a surface, and is given by
\begin{equation}
\sigma_{\text{surf}} = \vec{P}_{\text{bulk}} \cdot \vec{n}
\label{surface_charge}
\end{equation}
where $\vec{n}$ is the unit vector along the surface normal. A bulk polarization can occur of course in ferroelectric materials due to their {\it spontaneous polarization}. It can also occur in centrosymmetric crystals in which the ions form charged layers; we refer to this latter contribution as the {\it layer polarization}. 

Many mechanisms are known for screening bound surface charge, ranging from metal electrodes~\cite{Fong_et_al:2006, Sai/Kolpak/Rappe:2005}, to formation of charged defects~\cite{rappe_surface_str_switching, highland_prl_2011, gao_natcomm_2017, chrisholm_prl_2010, levchenko_prl_2008} or adsorption of charged species~\cite{setvin_science_2018, Strkalj_et_al:2019, Gattinoni_pnas_2020, garrity_prb_2013}, and even loss or reorientation of ferroelectric polarization~\cite{Junquera/Ghosez:2003, ishikawa_jjap_1996,Mundy_et_al:2020}. Note that it is not possible to screen the bound surface charge by surface relaxation alone, without the addition or removal of charged species.

In this work we explore the special case of multiferroic perovskite-structure bismuth ferrite, BiFeO$_3$, which has both a spontaneous polarization from its ferroelectric distortion and a layer polarization from its ionic charges. BiFeO$_3$ is particularly unusual because the size of its spontaneous ferroelectric polarization in the common [001] and [111] growth directions is close to the size of the layer polarization from the charged ionic layers in flat (100) and (111) planes. As a result, for certain choices of polarization orientation and surface termination, the spontaneous and layer polarizations self-compensate, leading to uncharged surfaces that are stable without external screening mechanisms.

We begin by reviewing in the next section (Sec.~\ref{sec:mod_theory_pol}) two key results of the modern theory of polarization --- the multivaluedness of the polarization lattice and the concept of the polarization quantum --- that are key to this work. We then briefly review compensating mechanisms at the surfaces of  ferroelectric materials with charge-neutral layers (Sec.~\ref{sec:charge_neutral}). Next, we discuss centrosymmetric materials with charged layers (Sec.~\ref{sec:non_polar}), and show how the layer polarization associated with charged ionic layers is conveniently described within the modern theory of polarization. We then  combine the concepts of layer polarization and spontaneous polarization in the example of bismuth ferrite (Sec.~\ref{sec:charged_layers}), discussing in turn its interface with metallic electrodes, and with insulators with different layer polarizations. 

\section{Reminder of key results from the modern theory of polarization}
\label{sec:mod_theory_pol}

We begin with a reminder of a fundamental result of the modern theory of polarization \cite{King-Smith/Vanderbilt:1994}, that the polarization, $\vec{P}$, of a bulk periodic solid, is not a single number but rather a {\it lattice} of values, separated by the polarization quantum, $\vec{P}_q = \frac{e\vec{R}}{V}$. The polarization quantum corresponds to the change in polarization on moving an electronic charge $e$ by a lattice vector $\vec{R}$ ($V$ is the unit cell volume), which changes the polarization by an amount $\vec{P}_q$, but does not change the physical system. (For a more extensive introductory discussion, see Ref.~\onlinecite{Spaldin:2012}.)

A consequence of this property is that the polarization lattice of a centrosymmetric crystal, which must also be centrosymmetric by symmetry, can take one of two sets of values, 
\begin{eqnarray}
\vec{P} & = & 0 + n \vec{P}_q \quad , \quad \text{or} \label{Eqn:P0}\\
\vec{P} & = & \frac{\vec{P}_q}{2} + n \vec{P}_q \quad ,
\label{Eqn:Phalf}
\end{eqnarray}
where $n$ is any integer. All insulating, centrosymmetric, periodic solids can be classified as belonging to one of these classes, which we will refer to as having ``zero-containing'' (Eq.~\ref{Eqn:P0}) or ``half-quantum-containing'' (Eq.~\ref{Eqn:Phalf}) polarization lattices. (In principle, it is possible for a centrosymmetric material to contain zero for one component of its polarization and a half quantum along another component, although we do not know of an example). Examples of half-quantum-containing centrosymmetric crystals are the high-symmetry paraelectric phase  of BiFeO$_3$, and the wide-band-gap insulating perovskite, LaAlO$_3$. Note also that these materials have charged ionic layers perpendicular to their usual [001] growth direction; in section ~\ref{sec:non_polar} we will show that these properties are formally connected. SrTiO$_3$ is an example of the zero-containing polarization lattice type; correspondingly charge-neutral layers, such as SrO and TiO$_2$ (001) planes, can be readily identified. 

While the fact that a centrosymmetric crystal can have a polarization lattice that does not contain zero is somewhat unintuitive, it is reconciled by a second important result of the modern theory of polarization: {\it Differences} in polarization, defined as the change in polarization along a given branch of the polarization lattice as the system is modified along an insulating pathway, are single-valued. As a result, the spontaneous polarization, which is the difference in polarization between the ferroelectric structure and its high-symmetry centrosymmetric counterpart, is single-valued.  Since only polarization differences (for example when a ferroelectric is switched between domains, or heated above its Curie temperature) are experimentally accessible, the theory is consistent with experimental reality. Indeed, when discussing the properties of infinite, bulk periodic crystals, the question of whether the polarization lattice contains zero or a half quantum is not generally relevant. 

At surfaces and interfaces, however, the question of the origin of the polarization lattice should not be disregarded. The bulk polarization,  $\vec{P}_{\text{bulk}}$, which gives rise to the bound charge at the surface of a crystal (Eqn.~\ref{surface_charge}) contains the contributions from both the half-polarization quantum (layer charges) {\it and} the spontaneous polarization \cite{Vanderbilt/King-Smith:1993}. Considering only the spontaneous polarization in evaluating the bound charge on the surface of a material will yield an incorrect result for centrosymmetric crystals with half-quantum-containing polarization lattices, as well as for ferroelectrics whose paraelectric reference structures' polarization lattices contain the half quantum. 

To illustrate these concepts in this work, we use ABO$_3$ perovskite-structure oxides in the usual pseudo-cubic [001] growth orientation. (For the generalization to other surface planes see Ref.~\onlinecite{Stengel:2011}). In this orientation, the layers have alternating AO / BO$_2$ chemistry, and the planar surfaces are formed from either entirely AO or entirely BO$_2$ layers. Since oxygen has formal charge -2, the (001) layers in so-called II-IV perovksite oxides (in which the A-site cation is divalent and the B-site cation has formal charge +4) are charge neutral. In III-III perovskite oxides (in which both A- and B-site cations have formal charge +3), the AO layer has charge +1 and the BO$_2$ layer has charge -1. 

As example II-IV ferroelectric materials with charge-neutral layers we choose PbTiO$_3$ and BaTiO$_3$; the polar discontinuity at their surfaces (that is their interface with the vacuum), and hence their bound surface charge, derives entirely from their spontaneous polarization. To illustrate the two possible behaviors of centrosymmetric ionic insulators, we choose SrTiO$_3$, with its neutral layers and zero-containing polarization lattice, and LaAlO$_3$, whose layers are charged, and whose polarization lattice contains the half quantum. The polar discontinuity between these two materials at their (001) interface has been of particular recent interest \cite{Ohtomo/Hwang:2004,Nakagawa/Hwang/Muller:2006,Reyren_et_al:2007}. The main part of this paper combines the concepts developed for both of these pairs of example materials to treat the case of BiFeO$_3$, with its combined spontaneous and layer polarizations.

\section{Surface effects in ferroelectric materials with zero layer polarization}
\label{sec:charge_neutral}

\begin{figure}[ht]
    \centering
    \includegraphics[scale=0.35, trim={0.5cm 0cm 2.5cm 0cm}, clip]{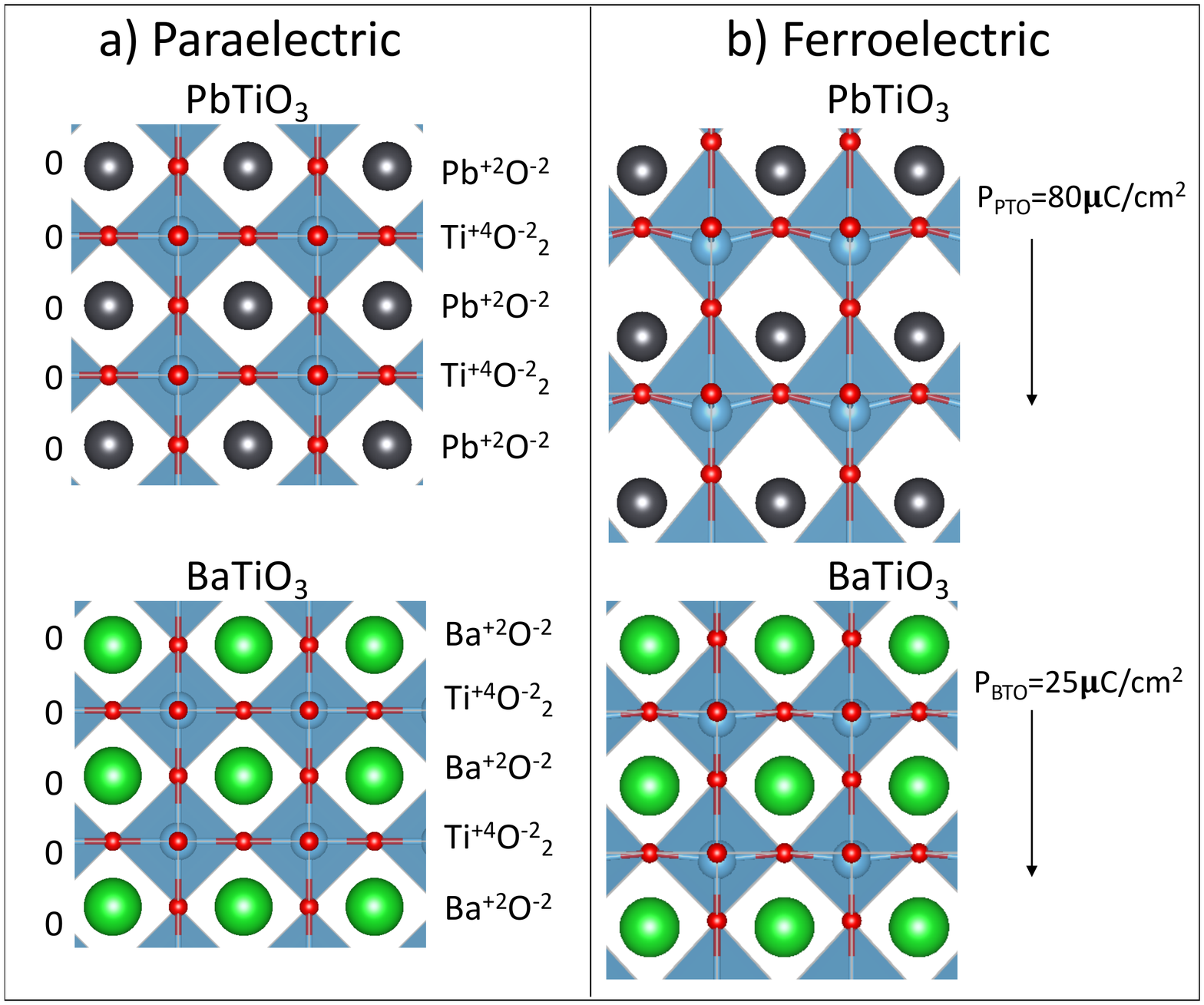}
    \caption{Crystal structure and (001) layer charges of PbTiO$_3$ (top) and BaTiO$_3$ (bottom). a) Arrangement of ions in the high-symmetry centrosymmetric reference structures. On the left of the crystal structure is the total charge for each layer, on the right the composition of each layer. b) Arrangement of ions in the ferroelectric structure. The arrows indicate the direction of the ferroelectric polarization. Pb is in black, Ti in blue, Ba in green and O in red.
    \label{fig:pto_bto}}
\end{figure}

\begin{figure}[ht]
    \centering
    \includegraphics[scale=0.5]{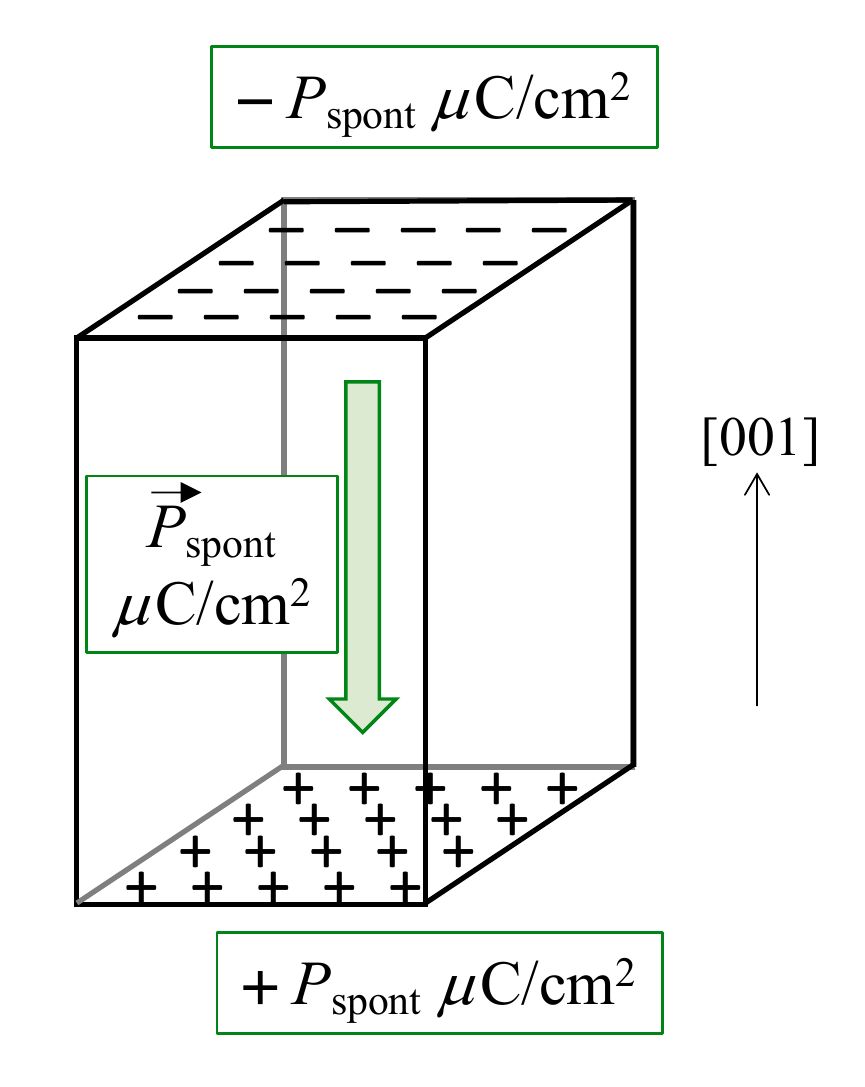}
    \caption{In ferroelectric materials with uncharged layers, such as BaTiO$_3$ and PbTiO$_3$, the sign and magnitude of the surface charge are determined by the spontaneous polarization, $\vec{P}_{\text{spont}}$. 
    \label{fig:Pspont}}
\end{figure}

We begin with a brief review of relevant results for the surface properties of the prototypical II-IV perovskite ferroelectrics, PbTiO$_3$ and BaTiO$_3$. In both cases, when the ions have the centrosymmetric arrangement of the high-symmetry reference phase, then the polarization lattice contains zero, and is given by Eqn.~\ref{Eqn:P0}. Correspondingly, the (001) layers have zero formal charges (Fig.~\ref{fig:pto_bto}). As a result, the only contribution to the $\vec{P}_{\text{bulk}}$ of Eqn.~\ref{surface_charge}, and hence to the surface charge, is from the spontaneous polarization. In PbTiO$_3$ the spontaneous polarization in the ferroelectric phase is oriented along a Cartesian axis ([001] in Fig.~\ref{fig:pto_bto}) and has the value $\sim$80 $\mu$C cm$^{-2}$; in BaTiO$_3$ there are a series of phase transitions with the polarization (of magnitude $\sim$25 $\mu$C cm$^{-2}$) reorienting from the cubic [111] to [011] to [001] directions as the temperature is lowered; we show the low-temperature [001] case in Fig.~\ref{fig:pto_bto}b. The bound surface charge on a (001) surface  (Fig.~\ref{fig:Pspont}) is then given trivially by the component of the spontaneous polarization perpendicular to the surface (80 or 25 $\mu$C cm$^{-2}$ for PbTiO$_3$ and BaTiO$_3$ respectively). It is negative (positive) on the upper surface for downward (upward) pointing polarization, and does not depend on the choice of layer (AO or BO$_2$) termination. In all cases, the uncompensated (001) surface is electrostatically unstable when the ferroelectric polarization is along the [001] axis. 

\begin{figure}[ht]
    \centering
    \includegraphics[scale=0.5, trim={1cm 9.5cm 14cm 0.5cm}, clip]{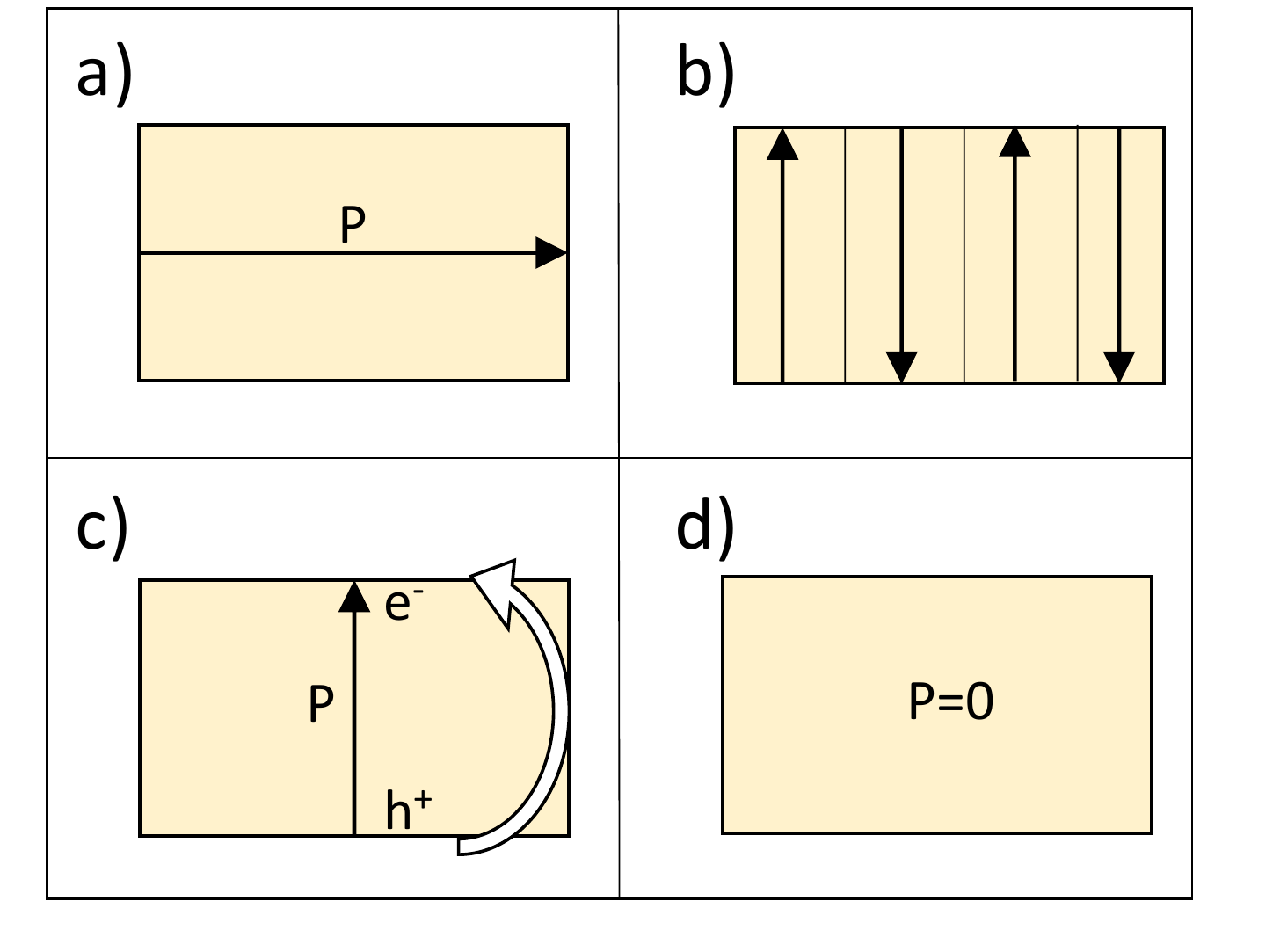}
    \caption{Effects of the depolarizing field on a ferroelectric thin film. a) In-plane polarization; b) Formation of domains; c) Accumulation of interfacial charge, for example by electron-hole excitation across the gap; d) Phase transition to a non-polar phase. 
    \label{fig:suppression}}
\end{figure}

The electrostatic instability associated with the spontaneous polarization in ferroelectrics is often discussed in terms of a depolarizing field, which acts in the opposite direction to the polarization to suppress the ferroelectricity in thin films. For a film with no compensation from external species, four main responses are known, as sketched in Fig.~\ref{fig:suppression}. Polarization reorientation into the plane of the film (Fig.~\ref{fig:suppression}a) completely eliminates the bound surface charge, and so is energetically favorable if it is not prohibited by, for example, strain effects~\cite{sarott_apl_2020}. If in-plane rotation of the polarization is unfavorable, the formation of domains can occur. The example of small domains of opposite orientation \cite{Fong_et_al:2004, Lubk-PhysRevLett:2012, Yang-NatNanotechnol:2010} shown in Fig.~\ref{fig:suppression}b reduces the overall charge on each surface; other more exotic textures such as polar skyrmions have also been reported~\cite{stachiotti_prl_2011,Yadav_et_al:2016,  damodaran_natmat_2017, Das_et_al:2019}. Screening surface charges can in principle be generated by electron-hole excitation across the band gap~\cite{Stengel:2011}  (Fig.~\ref{fig:suppression}c), although, since band gaps are typically of the order of an eV in ferroelectrics this is energetically expensive. Finally, complete suppression of the polarization (Fig.~\ref{fig:suppression}d) can occur, usually manifesting as a critical thickness of the paraelectric reference structure before the ferroelectric phase emerges \cite{Fong_et_al:2004, Stengel-PhysRevB:2012, DeLuca-NatCommun:2017}.

Screening can also occur from extrinsic factors.
When metallic electrodes are present, then carriers from the metal can provide compensating external surface charge to screen the polarization discontinuity.
Unless the screening is completely effective, however, the magnitude of the polarization and the ferroelectric Curie temperature, $T_c$, tend to be reduced from their bulk values~\cite{Junquera/Ghosez:2003, Dawber_et_al:2003, Kim_ApplPhysLett:2003, Sai/Kolpak/Rappe:2005, Tenne_et_al:2009} and there still tends to be a critical thickness below which the paraelectric phase is stable~\cite{Junquera/Ghosez:2003, puggioni_jap_2018}. In the absence of electrodes (or in combination with a bottom electrode), compensating charge can be provided by ions from the environment~\cite{levchenko_prl_2008}; this is the physics behind the well-known pyroelectric effect, in which the reduction in polarization on heating releases charged species from the surface~\cite{kakekhani_jmca_2016}. Recently, the reciprocal effect has been demonstrated, in which adsorption of adsorbates carrying a specific charge was shown to switch the ferroelectric polarization to achieve an electrostatically stable surface configuration~\cite{rappe_surface_str_switching, Gattinoni_pnas_2020, levchenko_prl_2008}.
Finally, we mention that the presence of charged ions in the growth chamber atmosphere has been exploited to enable growth of single-domain ultra-thin ferroelectric films of PbTiO$_3$ on SrRuO$_3$ through metalorganic chemical vapor deposition of PbTiO$_3$ on SrRuO$_3$ \cite{Fong_et_al:2006}. A charged atmosphere has even been shown to be more effective in screening the polarization than a top electrode for BaTiO$_3$ films grown on SrRuO$_3$ using pulsed laser deposition \cite{Strkalj_et_al:2019}.  

\section{Surface effects arising from the layer-charge polarization in non-ferroelectric materials}
\label{sec:non_polar}

\begin{figure}[ht]
    \centering
    \includegraphics[scale=0.4, trim={0cm 17cm 0cm 0cm}, clip]{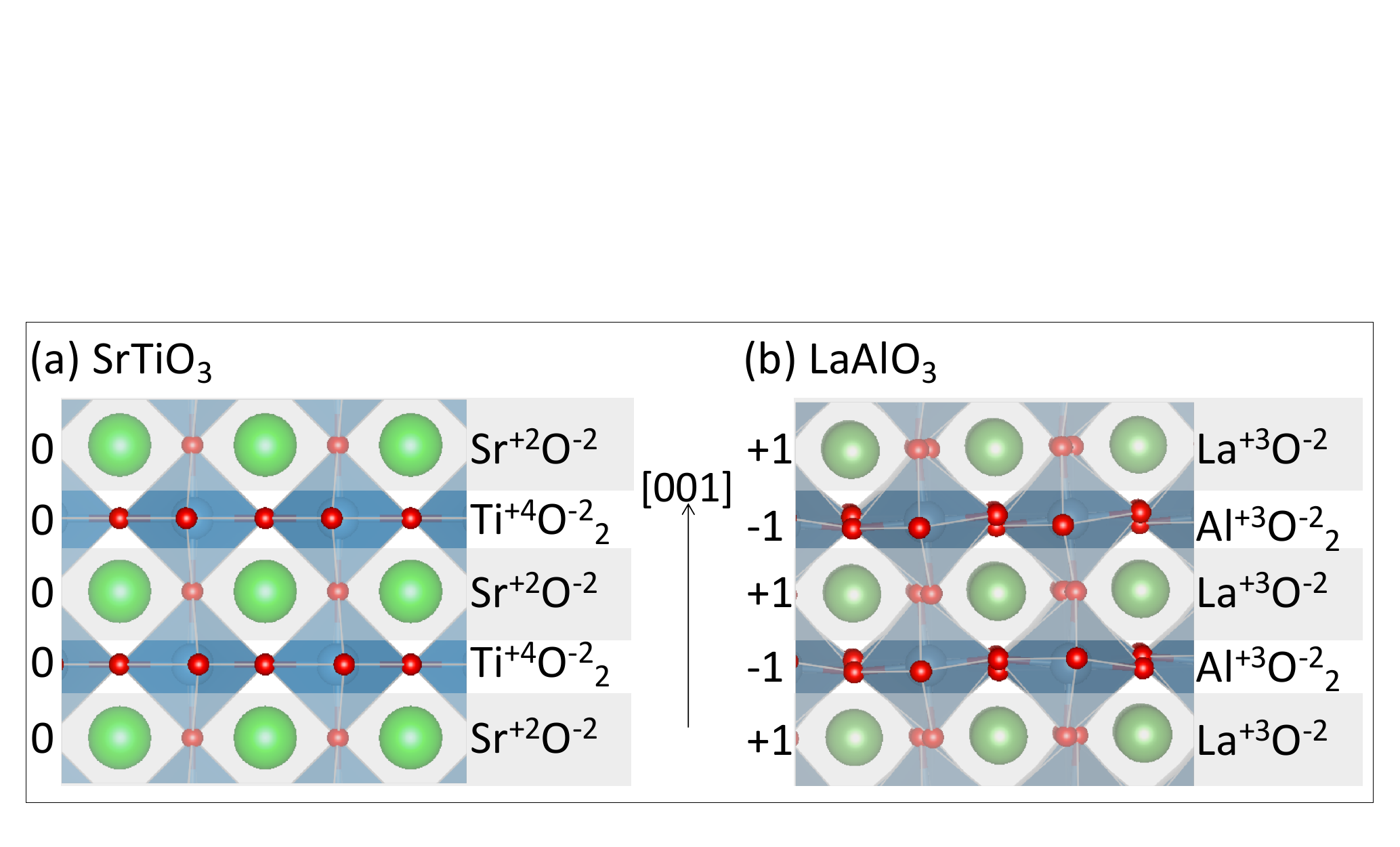}
    \caption{Crystal structure and (100) layer charges of a) SrTiO$_3$ and b) LaAlO$_3$. On the left of the crystal structure is the total charge for each layer, on the right the composition of each layer.
    \label{fig:sto_lao}}
\end{figure}

A pair of well-known centrosymmetric materials that illustrate the two cases of ``zero-containing'' or ``half-quantum-containing'' polarization lattices are the perovskite-structure oxides strontium titanate, SrTiO$_3$, and lanthanum aluminate, LaAlO$_3$, shown in Fig.~\ref{fig:sto_lao}. 
(Note that rotations of the oxygen octahedra, which lower the symmetry from the ideal cubic perovskite structure, occur in both materials; since these rotations preserve the center of inversion they do not change the polarization behavior and we do not consider them here.) We discuss next how the different polarization lattice types correspond to their different layer charges and result in different surface charges, focussing on the (001) surface for conciseness. For a more comprehensive discussion we direct the reader to Ref.~\onlinecite{Stengel:2011}.

\begin{figure}[ht]
    \centering
    \includegraphics[scale=0.4, trim={0cm 10cm 0cm 0cm}, clip]{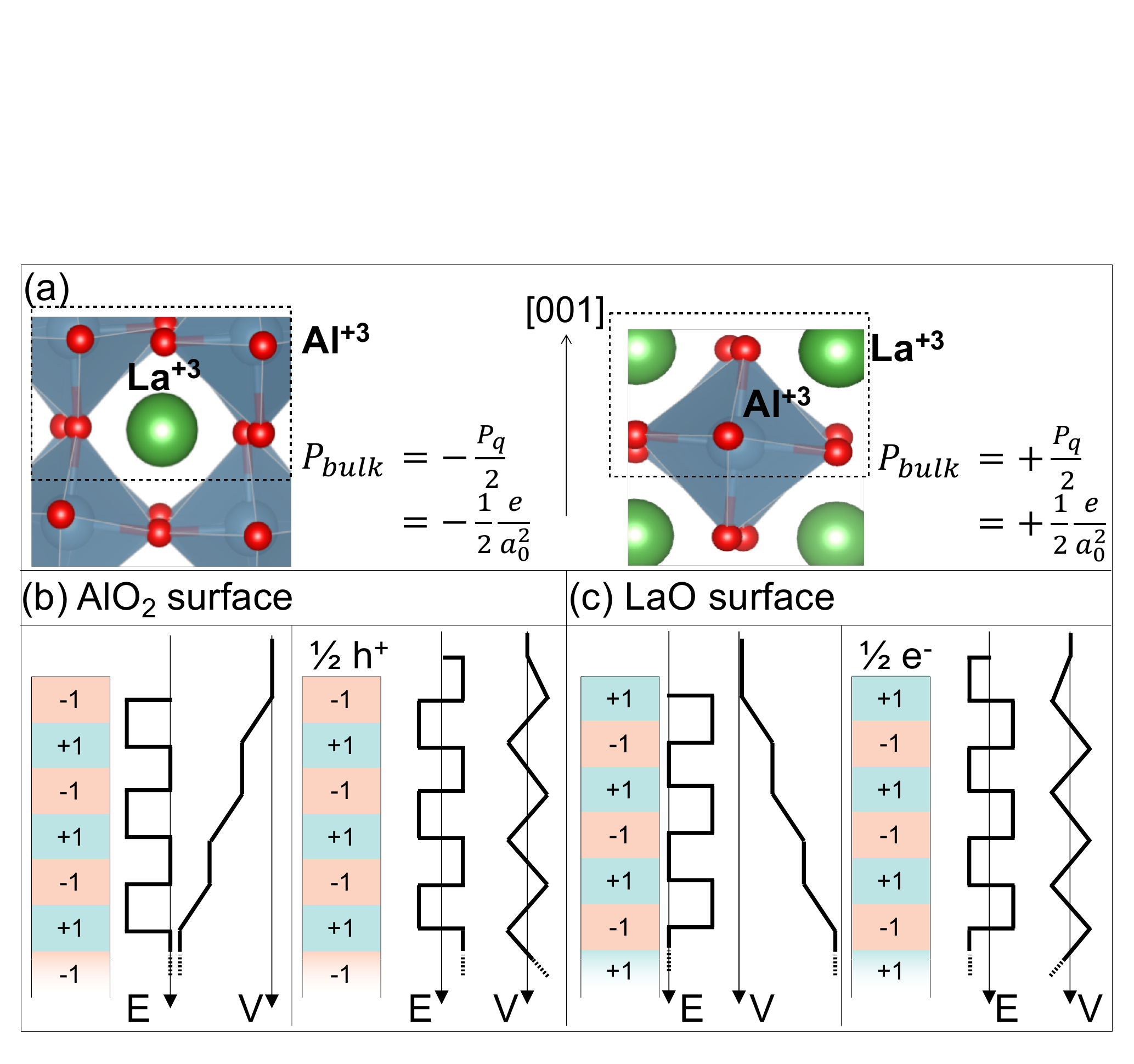}
    \caption{a) Unit cell of LaAlO$_3$ with AlO$_2$ (left) and LaO (right) termination. The dotted-line box indicates the basis used to calculate  $\vec{P}_{\mathrm{bulk}}$, and the surface termination is the topmost layer in the box. $\vec{P}_{bulk}$ in the [001] direction is of opposite sign for the two systems. b-c) Electric field E and potential V along the [001] direction for LaAlO$_3$ with b) AlO$_2$ and c) LaO termination. The system on the left in each panel has no compensation and so V diverges with thickness. The system on the right has a surface compensation of $1/2$ of the layer charge and therefore is stable.
    \label{fig:lao}}
\end{figure}

Since SrTiO$_3$ is a II-IV perovskite, and LaAlO$_3$ is a III-III perovskite, it is trivial to show, by calculating the polarization as $\vec{P} = \frac{1}{V} \Sigma_i Z_i \vec{r}_i$, that SrTiO$_3$ has the zero-containing polarization lattice of Eqn.~\ref{Eqn:P0}, and LaAlO$_3$ the half-quantum-containing form of Eqn.~\ref{Eqn:Phalf}. Here, $Z_i$ are the formal ionic charges and $\vec{r}_i$ their positions  within any choice of unit cell. (Note that rigorous calculation using the Berry phase formalism gives an identical result, and the use of the formal charges when calculating the lattice polarization of the paraelectric structure is formally correct~\cite{Stengel:2011}.) As a result, the (001) surface of SrTiO$_3$ has a charge of zero or $n \vec{P}_q$, whereas that of LaAlO$_3$ has a charge of $\frac{n\vec{P}_q}{2}$.

A convenient recipe was provided in Ref.~\onlinecite{Stengel:2011} to determine which branch of the polarization lattice (that is, which value of $n$) is the relevant one for a particular choice of surface plane and chemistry: Take the unit cell that tiles the semi-infinite slab containing the surface of interest, and calculate the dipole per unit volume for that unit cell. The answer is $\vec{P}_{\text{bulk}}$ with the appropriate choice of polarization lattice branch. For the (001) surface of SrTiO$_3$, both smooth surfaces (containing SrO or TiO$_2$) yield  $\vec{P}_{\text{bulk}}=0$ with this recipe. Therefore they have no bound charge and do not require any external charge compensation. For the (001) surface of LaAlO$_3$ (Fig.~\ref{fig:lao}a), the LaO surface has [001] polarization value $\vec{P}_{\text{bulk}} = + \frac{1}{2}\frac{|e|}{a_0^2} = +\frac{\vec{P}_q}{2}$, requiring compensation by a negative of charge of this size and the AlO$_2$ surface has $\vec{P}_{\text{bulk}} = - \frac{1}{2}\frac{|e|}{a_0^2} = - \frac{\vec{P}_q}{2}$, requiring compensation by the corresponding positive charge. ($a_0$ is the length of the pseudo-cubic unit cell, and the polarization quantum is $\vec{P}_q = \frac{|e|}{a_0^2}$.) The required compensation of 
half an electronic charge per simple cubic unit cell corresponds to the convenient value of 50 $\mu$C cm$^{-2}$ in the conventional units used in the ferroelectrics literature, taking a cubic lattice constant of 4 \AA\, which is slightly larger than the values for SrTiO$_3$ ($\sim$ 3.9 \AA\ ) and LaAlO$_3$ ($\sim$ 3.8 \AA\ ). 

An alternative picture that is intuitively appealing, although not as rigorously well-founded, is to decompose the materials into planes of ions and consider the net charges of these planes, as shown in Fig.~\ref{fig:sto_lao}. For the case of II-IV perovskites such as SrTiO$_3$, the (001) planes are alternately SrO and TiO$_2$, both of which are charge neutral. Therefore any planar (001) surface in a II-IV perovskite carries no net surface charge and so is stable. In III-III perovskites such as LaAlO$_3$, the (001) planes are alternately LaO and AlO$_2$ with charges $+1$~$e$ and $-1$~$e$ per unit cell respectively. 
And so, depending on the choice of termination, the surface has the corresponding positive or negative charge per surface unit cell, that is $\pm 100$ $\mu$C cm$^{-2}$. As illustrated in Fig.~\ref{fig:lao}b, the positive layer charge of the LaO surface requires a compensating negative charge of {\it half} the layer charge, that is $-0.5$ $e$ per unit cell or $- 50$ $\mu$C cm$^{-2}$, to prevent a divergence of the electrostatic potential and stabilize the surface. (A compensating charge {\it equal to} the surface charge just displaces the problem to a new terminating layer \cite{Hwang:2006}.) Likewise, the formally negatively charged AlO$_2$ surface (Fig.~\ref{fig:lao}c) requires a compensating positive charge of the same amount. Thus we reach the same conclusion as that derived from consideration of the bulk polarization. 

The implication of the different bulk polarization lattices of LaAlO$_3$ and SrTiO$_3$ for the {\it interface} between the two materials is profound: The polarization discontinuity between the two materials means that it is not possible to make a stoichiometric interface that is electrostatically stable~\cite{Stengel/Vanderbilt:2009}.
Specifically, an SrO / AlO$_2$ interface requires a compensating positive charge of magnitude half an electronic charge per unit cell, and the LaO / TiO$_2$ interface requires half an electronic charge per unit cell of negative charge. In the latter case, the extra electrons occupy the broad Ti $3d$-derived energy bands at the bottom of the valence band. The compensating electrons are therefore mobile and form an interfacial two-dimensional electron gas~\cite{Ohtomo/Hwang:2004}, a remarkable behavior for the interface of two robust band insulators. The electron gas has even been shown to be superconducting at low temperature \cite{Reyren_et_al:2007}.

Note that these considerations are not limited to III-III perovskites, but are relevant for the surfaces and interfaces of all centrosymmetric insulators that have a ``half-quantum containing'' polarization lattice.
Another example is provided by the I-V perovskites, such as KTaO$_3$, where surface reconstructions~\cite{setvin_science_2018} and surface and interface 2D electron gases~\cite{santander_prb_2012, king_prl_2012} have been observed.

\section{Surfaces of ferroelectric materials with charged layers -- the interplay of layer charge and spontaneous polarization in bismuth ferrite}
\label{sec:charged_layers}

Next we turn to the case of ferroelectric materials whose polarization lattice in their high-symmetry centrosymmetric prototype structure contains the half-polarization quantum. We choose the example of the III-III ferroelectric perovskite BiFeO$_3$, which, as mentioned in the introduction, combines a half-quantum-containing polarization lattice in its centrosymmetric reference structure, with a spontaneous polarization of almost exactly 50 $\mu$C cm$^{-2}$ in the [001] direction. In particular, we will explore the consequences of the accidental layer- and spontaneous polarization-surface charge compensation on the stability of thin films and heterostructures of BiFeO$_3$. 

\begin{figure}[ht]
    \centering
    \includegraphics[scale=0.3]{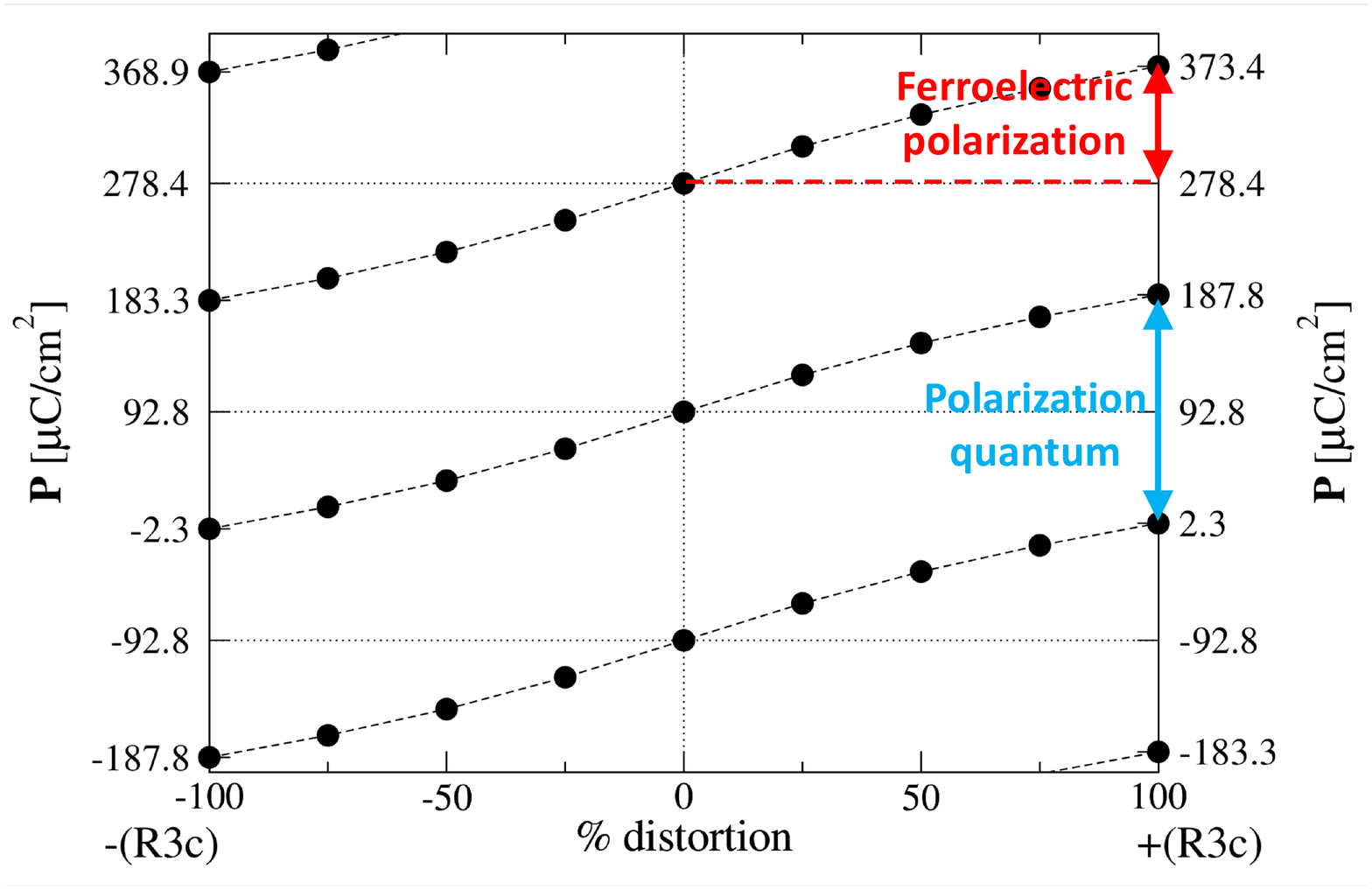}
    \caption{Polarization in the [111] direction for BiFeO$_3$, calculated using the LSDA$+U$ and Berry phase methods within density functional theory in Ref. \onlinecite{Neaton_et_al:2005}. $\pm 100$\% distortion corresponds to the ground-state $R3c$ structures of opposite polarization; 0\% distortion corresponds to the ideal cubic perovskite structure. Three full branches of the polarization lattice are shown. The central branch illustrates that, starting from a centrosymmetric polarization of 92.8 $\mu$C cm$^{-2} (= \frac{\vec{P}_q}{2})$ and introducing a negative spontaneous polarization yields a bulk polarization close to zero (in fact -2.3 $\mu$C cm$^{-2}$) and a correspondingly small surface charge. Introducing a positive spontaneous polarization in this branch results in a very large bulk polarization (187.8 $\mu$C cm$^{-2}$) and an unfavorably large surface charge. The lower branch illustrates the opposite scenario. Reproduced from Ref. \onlinecite{Neaton_et_al:2005}. Copyright 2005 by the American Physical Society.
    \label{fig:BiFeO$_3$_path}}
\end{figure}

The ground state of bulk BiFeO$_3$ has the $R3c$ structure, which is reached from the prototypical cubic perovskite structure by alternating rotations of the oxygen octahedra around the [111] axis, combined with opposite displacements of anions and cations along the [111] direction. The latter results in a large spontaneous polarization of magnitude $\sim$ 90 $\mu$C cm$^{-2}$ oriented along [111]. In Fig.~\ref{fig:BiFeO$_3$_path} we show the evolution of the polarization (calculated using the Berry phase approach in Ref.~\onlinecite{Neaton_et_al:2005}) as a function of the amplitude of the ferroelectric distortion from the high-symmetry reference structure (0\% distortion) to the ground-state ferroelectric structure (100\% distortion), for several branches of the polarization lattice.  The spontaneous ferroelectric polarization along [111] is highlighted in red.
Interestingly, and completely coincidentally, the value of the spontaneous polarization is very close to half of the polarization quantum of $\sim$ 180 $\mu$C cm$^{-2}$ along the [111] direction (highlighted in blue in Fig.~\ref{fig:BiFeO$_3$_path}) for BiFeO$_3$. Since the polarization lattice for the centrosymmetric reference structure is of the half-quantum type, we see that there are two combinations of the centrosymmetric layer polarization and the spontaneous polarization ($+\frac{\vec{P}_q}{2}$ with $P_{\text{spont}} = -90$ $\mu$C cm$^{-2}$, and $-\frac{\vec{P}_q}{2}$ with $P_{\text{spont}} = +90$ $\mu$C cm$^{-2}$) that combine to give a bulk polarization value, $\vec{P}_{\text{bulk}}$, in the ferroelectric structure that is very close to zero (in fact $\pm 2.3$ $ \mu$C cm$^{-2}$ in Fig.~\ref{fig:BiFeO$_3$_path}).

This in turn leads to a cancellation of the bound surface charge, $\sigma_{\text{surf}} \approx 0$. 
A consequence of this cancellation, therefore, is that free-standing thin films of BiFeO$_3$ are  electrostatically stable for one choice of polarization for each surface; this has been referred to as the ``happy'' configuration in the literature~\cite{efe_jcp_2020}.

\begin{figure}[ht]
    \centering
    \includegraphics[scale=0.4, trim={0cm 6cm 12cm 0cm}, clip]{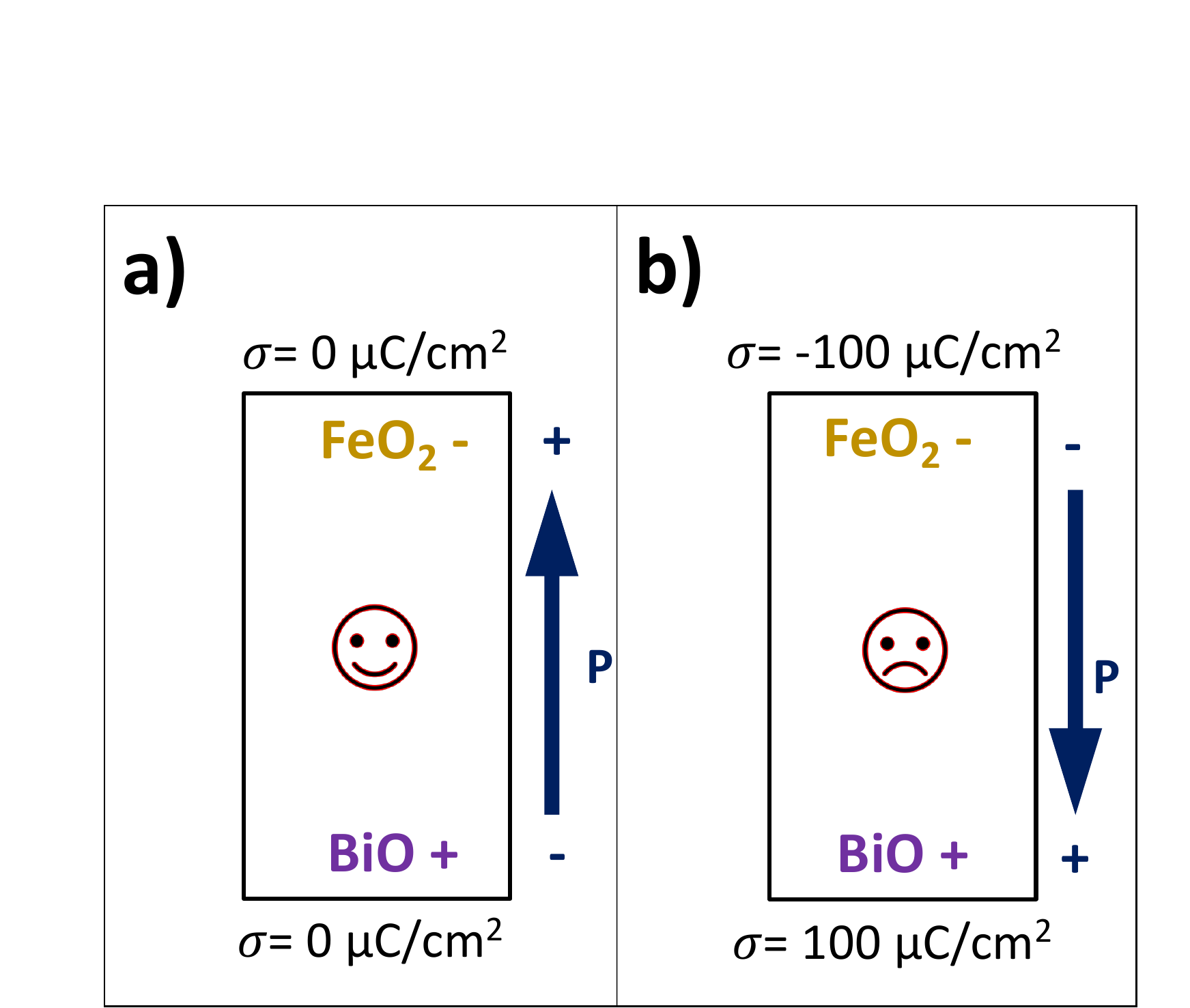}
    \caption{Combinations of ferroelectric polarization direction and surface termination leading to a) stable (happy) and b) unstable (unhappy) BiFeO$_3$ (001) surfaces. In a), the spontaneous polarization, $P_{\text{spont}} = 50$ $\mu$C cm$^{-2}$ is compensated by the  layer polarization. In b) the layer polarization adds to the spontaneous polarization $P_{\text{spont}} = -50$ $\mu$C cm$^{-2}$ to give a surface charge of $ \pm 100$ $\mu$C cm$^{-2}$.
    \label{fig:BiFeO$_3$_surface}}
\end{figure}

In Fig.~\ref{fig:BiFeO$_3$_surface} we illustrate this with a cartoon of a free-standing BiFeO$_3$ slab in the commonly grown [001] orientation. The projection of the [111]-oriented ferroelectric polarization into the [001] direction results in a spontaneous [001] polarization of $\sim \pm 50$ $\mu$C cm$^{-2}$. As we saw in the case of LaAlO$_3$ in Section~\ref{sec:non_polar}, the unit cell corresponding to the BO$_2$ (FeO$_2$ in this case) surface selects for the branch on the centrosymmetric polarization lattice with value $-\frac{1}{2} \frac{e}{a_0^2} = -50$ $\mu$C cm$^{-2}$. Therefore an FeO$_2$ surface with a positive (i.e. pointing towards it, or upwards in Fig.~\ref{fig:BiFeO$_3$_surface}a) value of spontaneous polarization has zero surface charge and is stable; conversely the AO (BiO in this case) surface selects for the  $+\frac{1}{2} \frac{e}{a_0^2} = +50$  $\mu$C cm$^{-2}$ half quantum, and requires a negative (i.e. pointing away from it) polarization to ensure stability. Note that the opposite combinations are twice as unfavorable (``unhappy'') as they would be in a II-IV perovskite with the same magnitude of spontaneous polarization but uncharged layers (Fig.~\ref{fig:BiFeO$_3$_surface}b), since they would have a surface charge of $\pm 100$ $\mu$C cm$^{-2}$.
In the alternative charged-layers picture, the centrosymmetric BiFeO$_3$ is composed of alternating (001) layers of positively charged BiO (+1 $e$ per unit cell or $\sim + 100 \mu$C cm$^{-2}$) and negatively charged FeO$_2$ (-1 $e$ per unit cell or $\sim - 100 \mu$C cm$^{-2}$). The appropriately oriented spontaneous polarization of magnitude $50$ $ \mu$C cm$^{-2}$ then provides the required  compensating surface charge of half that amount. 

Next, we discuss the consequences of this layer and spontaneous polarization cancellation, examining examples from the literature as well as presenting new results of behaviors that are caused by the happiness or unhappiness of BiFeO$_3$ surfaces and interfaces. We consider three scenarios: first, BiFeO$_3$ on a metallic substrate, followed by interfaces with centrosymmetric II-IV then III-III insulators.

\subsection{Interaction of BiFeO$_3$ thin films with a metal substrate}

We begin with the case of BiFeO$_3$ films grown on  substrates that are metallic, and therefore provide good screening of any interfacial charge at the bottom interface. We expect, therefore, that the orientation of the spontaneous polarization will be determined by the nature of the top surface with the vacuum.

We take the examples of BiFeO$_3$ on two metallic oxides, La$_{0.7}$Sr$_{0.3}$MnO$_3$ (LSMO) and SrRuO$_3$. Heterostructures of these combinations were grown and characterized in Ref.~\onlinecite{Yu_pnas_2012}, and we begin by analyzing the results of that work in the context of the surface electrostatics introduced above. 

In Ref.~\onlinecite{Yu_pnas_2012}, the SrRuO$_3$ substrate was terminated with an SrO layer, and so the bismuth ferrite film, which grows in complete BiFeO$_3$ unit cells, began with an FeO$_2$ layer and ended with a BiO surface.
As expected, the polarization spontaneously adopted the down orientation, corresponding to zero surface charge. 
An FeO$_2$ surface was achieved for BiFeO$_3$ on SrRuO$_3$ by inserting a monolayer of TiO$_2$ at the interface so that the BiFeO$_3$ film began with a BiO layer. 
This caused a spontaneous upwards polarization in the BiFeO$_3$, again corresponding to the zero surface-charge configuration as expected.

\begin{figure}[ht]
\centering
 \includegraphics[scale=1.0]{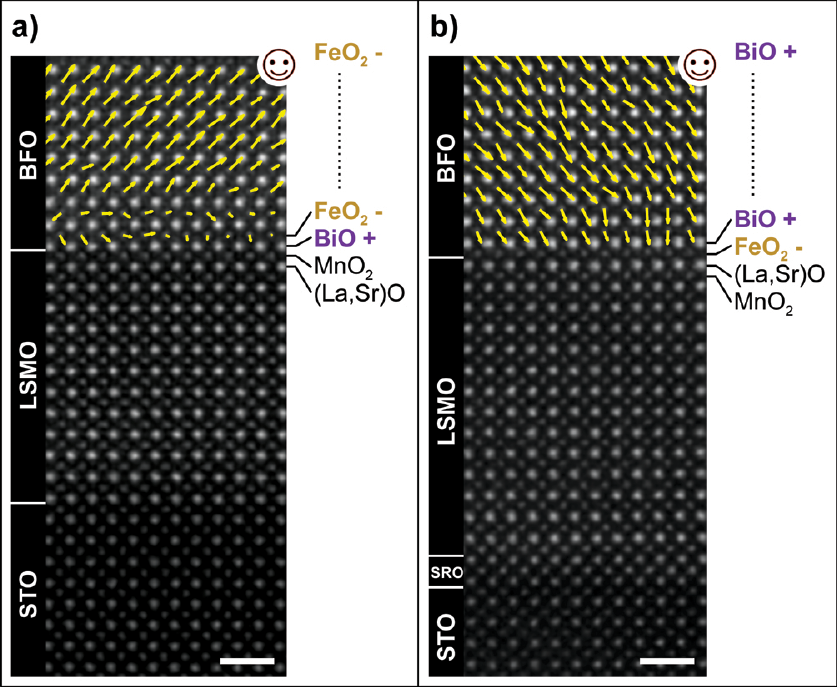}
 \caption{Cross-sectional HAADF-STEM images of two distinct BiFeO$_3$/La$_{0.7}$Sr$_{0.3}$MnO$_3$ heterointerfaces with overlaid vector maps showing the polarization in the BiFeO$_3$ layers. a) The sample with the MnO$_2$-terminated (La$_{0.7}$Sr$_{0.3}$O-MnO$_2$-BiO-FeO$_2$) interface develops a spontaneous up-pointing polarization ($\vec{P}$ pointing towards the FeO$_2$ surface). b) The sample with the La$_{0.7}$Sr$_{0.3}$O-terminated (MnO$_2$-La$_{0.7}$Sr$_{0.3}$O-FeO$_2$-BiO) interface develops a spontaneous down-pointing polarization ($\vec{P}$ pointing away the BiO surface). The scale bar is 1 nm.}
 \label{fig:BiFeO$_3$onLSMO}
\end{figure}

Growth of BiFeO$_3$ on LSMO shows a similar behavior. In  Fig.~\ref{fig:BiFeO$_3$onLSMO} we show two high-angle annular dark-field  scanning transmission electron microscopy (HAADF-STEM) images of the BiFeO$_3$ on LSMO heterostructures grown in Ref.~\onlinecite{Yu_pnas_2012}. (For additional details about the thin film growth see section~\ref{sec:Methods} and Ref.~\onlinecite{Yu_pnas_2012}.) In panel a the LSMO is terminated with MnO$_2$, so the BiFeO$_3$ layer has an FeO$_2$ surface, while in panel b the LSMO is (La,Sr)O terminated, so the bismuth ferrite starts with a FeO$_2$ layer and has a BiO surface. We have overlaid arrows, which are vector maps indicating the local polarization extracted from the measured atomic positions, in the BiFeO$_3$ layers.
Again, as expected from the surface electrostatics, we see that case (a) develops a spontaneous up-pointing polarization ($\vec{P}$ pointing towards the FeO$_2$ surface) and case (b) a spontaneous down-pointing ($\vec{P}$ pointing away from the BiO surface).
In all four scenarios switching of the polarization was achieved using a tip in a piezoforce geometry, but with considerable exchange bias favoring the spontaneous orientation. 

In our discussion so far, we have assumed that the 
LSMO and SrRuO$_3$ substrates behave like ideal metals, and have disregarded the fact that their constituent ions have different formal layer charges ($\pm \sim 0.7$ \emph{e} for LSMO at 0.3 Sr concentration, and neutral for SrRuO$_3$). The different formal layer charge discontinuities could play a role if the metallic screening is incomplete. Indeed, it is known that ionic relaxation is an important contributor to surface-charge screening in oxide electrodes \cite{Stengel/Spaldin_Nature:2006}, and the use of polar metals as electrodes has been proposed as a route to overcoming the critical thickness in ferroelectric capacitors~\cite{puggioni_jap_2018}. 
To investigate the role of the formal layer charges in metallic oxide electrodes, we next perform density functional calculations of [001]-oriented BiFeO$_3$/SrRuO$_3$ superlattices in both the happy (Fig.~\ref{fig:BiFeO$_3$_surface}a) and sad (Fig.~\ref{fig:BiFeO$_3$_surface}b) interfacial orientations (for details see the methods in Section~\ref{sec:Methods}).

\begin{figure}[ht]
    \centering
    \includegraphics[scale=0.47, trim={0cm 0cm 0cm 0cm}, clip]{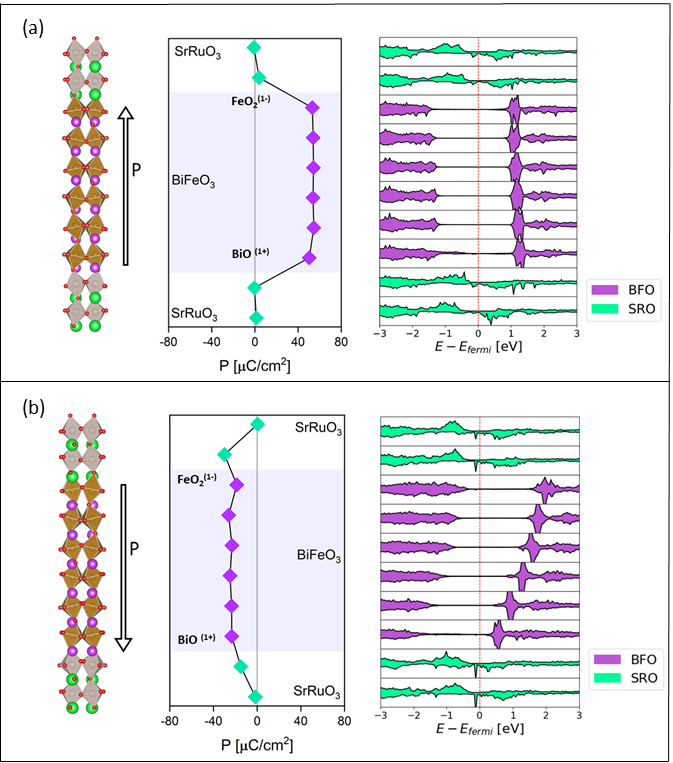}
    \caption{Calculated structures (left), layer by layer polarizations (middle) and layer density of states (right) for the two BiFeO$_3$/SrRuO$_3$ heterostructures studied in this work. Panel a) shows the happy BiFeO$_3$ slab configuration; panel b) the unhappy. Green symbols and shading indicate SrRuO$_3$; BiFeO$_3$ is shown in purple.
    \label{fig:BFOSROsummary}  }
\end{figure}

We constructed two superlattices each containing six layers of BiFeO$_3$ and four layers of SrRuO$_3$, with one SrO/FeO$_2$ and one BiO/RuO$_2$ interface, see Fig.~\ref{fig:BFOSROsummary}. The  two supercells had opposite orientation of the BiFeO$_3$ ferroelectric polarization, such that one system had self-compensating (Fig.~\ref{fig:BFOSROsummary}a) and the other charged interfaces (Fig.~\ref{fig:BFOSROsummary}b). In both cases, the entire heterostructure adopts the a$^-$a$^-$c$^-$ tilt pattern of BiFeO$_3$ (Fig.~\ref{fig:BFOSROsummary}, left panels), and both materials maintain their bulk magnetic orderings (G-type antiferromagnetic for BiFeO$_3$ and ferromagnetic for SrRuO$_3$). 
As expected, both from our electrostatic arguments and from the experimentally observed exchange bias~\cite{Yu_pnas_2012}, the happy system is energetically the most stable, $\sim$ 2 eV per supercell lower in energy than the unhappy system. 

The calculated structures, layer-by-layer polarizations and layer-resolved densities of states are shown in Fig.~\ref{fig:BFOSROsummary}. Note that, in both the ``happy'' and ``unhappy'' systems, the SrRuO$_3$ layers are metallic (in green in the density of states graph on the right-hand-side on Fig.~\ref{fig:BFOSROsummary}), with a finite density of states at the Fermi energy in all layers, and the BiFeO$_3$ (in purple) is insulating, with the Fermi energy (shown as a vertical red line) lying in the gap. 
The layer-resolved densities of states for the happy system (right-most panel of Fig.~\ref{fig:BFOSROsummary}a) indicate that there is no band bending and hence no internal electric field in the happy system, consistent with the absence of surface charge in the happy BiFeO$_3$ slabs. The polarization (middle panel) has its full bulk value throughout the slab, and drops abruptly to zero in the first layer of the SrRuO$_3$. 

For the six-unit-cell BiFeO$_3$ heterostructures that we present here, we find that the unhappy system is metastable in our DFT calculations. (For thinner films the polarization orientation reverses and the structure relaxes to the happy system.)
We find, however, a suppressed layer polarization compared with the happy system, as visible in the middle graph of Fig.~\ref{fig:BFOSROsummary}b, as well as a pronounced shift in the BiFeO$_3$ band edges from layer to layer indicating a strong internal electric field resulting from the large uncompensated surface charges in the BiFeO$_3$ slab. 
In addition, the interfacial SrRuO$_3$ layers undergo a polar ionic distortion to further reduce the polar discontinuity, similar to that observed in Ref.~\citenum{puggioni_jap_2018} for the SrRuO$_3$/BaTiO$_3$ interface.

In summary, our calculations indicate that, even with metallic screening, the direction of polarization preferred  by the interplay between the lattice and spontaneous polarization of the ferroelectric layer is strongly preferred. While screening by the metal is able to stabilize the unhappy polarization orientation it is still energetically unfavorable. This behavior explains the strong electric-field exchange bias effects \cite{Maksymovych_et_al:2009}, as well as the highly asymmetric resistive switching \cite{Shuai_et_al:2011} found in BiFeO$_3$ capacitors.

As a final example from the literature, we choose the case of heterostructures between BiFeO$_3$ and the metallic lightly-doped II-IV magnetic insulator, CaMnO$_3$. Heterostructures of Ca$_{1–x}$Ce$_x$MnO$_3$/BiFeO$_3$ were grown on YAlO$_3$, whose small lattice constant causes a strongly compressive biaxial in-plane strain and correspondingly large out-of-plane lattice constant and polarization (see detailed discussion in Subsection C) ~\cite{marinova_nanolett_2015}. The ferroelectric polarization was determined to have an out-of-plane value of $\sim 100$ $\mu$C/cm$^{-2}$, and to point towards the BiFeO$_3$/CaMnO$_3$ interface, which was of the FeO$_2^-$ -- CaO type. While this is the least unhappy arrangement, in this case, because of the unusually large out-of-plane polarization, the FeO$_2^-$ layer provides only partial compensation of the bound surface charge. An additional electronic charge accumulation of $\sim 0.65$ electrons per unit cell area was found using electron energy loss spectroscopy (EELS) to accumulate in the near-interfacial Ca$_{1–x}$Ce$_x$MnO$_3$ layers.

\subsection{Interfaces of bismuth ferrite with centrosymmetric II/IV insulating perovksites}

\begin{figure}[ht]
\centering
 \includegraphics[width=0.5\textwidth, trim={0cm 0cm 7cm 0cm}, clip]{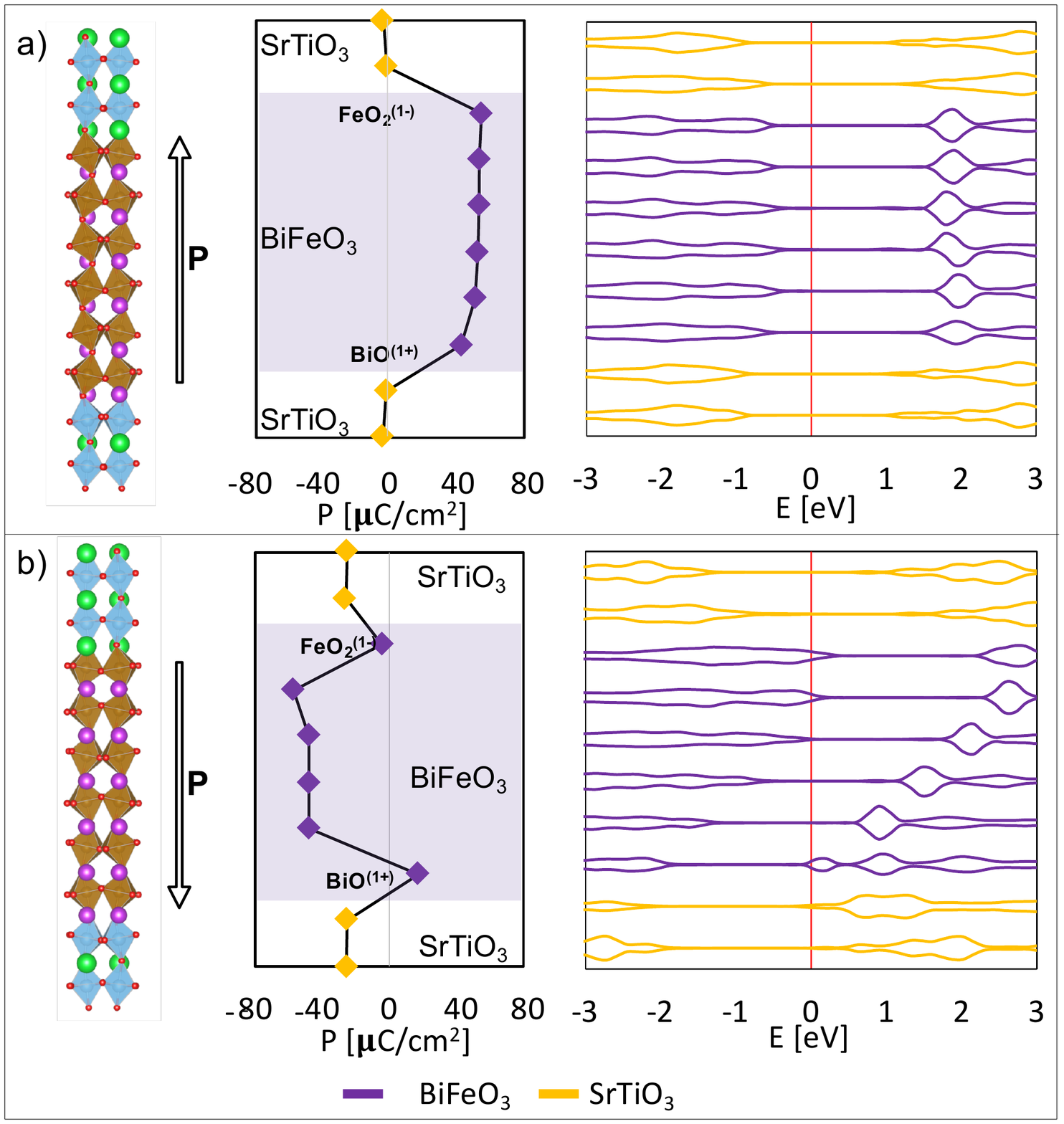}
 \caption{Polarization along the [001] direction of a (SrTiO$_3$)$_4$/(BiFeO$_3$)$_6$ heterostructure with one BiO/TiO$_2$ interface and one FeO$_2$/SrO interface. The polarization direction leading to a happy system is in panel a, and to an unhappy system in panel b. The systems are shown on the left, in the middle their corresponding unit cell-by-unit cell polarization and on the right the unit cell-by-unit cell local density of states. Orange data points correspond to SrTiO$_3$ and purple ones to BiFeO$_3$. In the atomic structure Sr is green, Ti blue, Bi purple, Fe brown and O red.}
 \label{fig:sto_bfo}
\end{figure}

Since the (100) surface of a centrosymmetric II/IV insulating perovskite such as SrTiO$_3$ has zero bound charge, we expect it to behave electrostatically similarly to the vacuum in its interface with BiFeO$_3$. 
That is, we expect that BiFeO$_3$ surfaces that are happy in free-standing slabs to form stable interfaces with SrTiO$_3$, with bulk-like ferroelectricity in the BiFeO$_3$ layer down to small thicknesses, whereas electrostatically unstable surfaces will have similarly unhappy interfaces with SrTiO$_3$ and their ferroelectric polarization will tend to reverse~\cite{efe_jcp_2020}. 
All the examples that we have been able to find of experimental BiFeO$_3$/SrTiO$_3$ superlattices and heterostructures have their as-grown polarization orientation  in the direction which compensates the layer polarization, consistent with this assumption~\cite{Bruyer_et_al:2015, Ranjith_et_al:2010, Yang_et_al:2013, Gao_et_al:2019}. Likewise, literature calculations of PbTiO$_3$/BiFeO$_3$ superlattices, with the polarization of the PbTiO$_3$ entirely in the plane of the superlattice~\cite{Yang_et_al:2012}, found a stable solution with BiFeO$_3$ in its happy configuration.

 To explore the details of the behavior, we perform DFT calculations for two [001] superlattices, each containing four layers of SrTiO$_3$ and six layers of BiFeO$_3$ with one SrO / FeO$_2$ and one BiO / TiO$_2$ interface, but with the BiFeO$_3$ polarization initialized to opposite orientations.
As expected, we find that the happy case is stable, with the layer polarization close to the bulk value throughout the film (Fig.~\ref{fig:sto_bfo}a, middle panel). 
As in the case of the interface with SrRuO$_3$, the unit cell-by-unit cell density of states indicates zero internal electric field in ferroelectric BiFeO$_3$ since the charge compensation between the spontaneous and layer polarization leads to zero surface charge.
In contrast, we are unable to stabilize the unhappy state in our DFT calculations unless we constrain the polarization orientation in the middle layers of the BiFeO$_3$ slab (Fig.~\ref{fig:sto_bfo}b).
The layer-by-layer density of states indicates a strong band bending due to the large internal electric field, and the formation of metallic layers by electron-hole excitation across the band gap at the interfaces. (Similar metallicity at the unhappy BiO / TiO$_2$ interface was seen in an earlier DFT calculation for a non-stoichiometric BiFeO$_3$ slab with both of its interfaces set to BiO sandwiched between two SrTiO$_3$ layers~\cite{Zhang_et_al:2011}.) Additionally, the SrTiO$_3$, which is an incipient ferroelectric and therefore readily polarizable, develops a polarization parallel to the spontaneous polarization of BiFeO$_3$ to reduce the polar discontinuity. 

\subsection{Interfaces of bismuth ferrite with insulating centrosymmetric III-III perovskites}

Finally, we consider the case of the interface between BiFeO$_3$ and a centrosymmetric insulating III-III perovskite, which contains $\frac{\vec{P}_q}{2}$ in its polarization lattice. 
Here the centrosymmetric contributions to the polarization lattice are similar in both materials (they will differ slightly if the lattice vectors and unit cell volumes are different) and so the interfacial polar discontinuity is given by the difference in spontaneous polarizations. 
If the second material is centrosymmetric, then the polar discontinuity is equal to the spontaneous polarization of BiFeO$_3$, similar to the case of the interfaces between the II-IV ferroelectrics PbTiO$_3$ or BaTiO$_3$ and vacuum discussed above. 

%
%

A number of different routes to avoiding the polarization discontinuity at the interface, of the types we summarized in Fig.~\ref{fig:suppression}, have been observed. An in-plane polarization (Fig.~\ref{fig:suppression}a) associated with an orthorhombic phase has been reported for BiFeO$_3$ on a NdScO$_3$ substrate, which also imparts a small biaxial tensile strain \cite{Yang_et_al:oBFO:2012}. Since many low-energy metastable non-polar, anti-polar and even antiferroelectric phases of BiFeO$_3$ are known ~\cite{Stengel/Iniguez:2015,Dieguez_et_al:2011}, the BiFeO$_3$ film can also lose its polarization entirely (Fig.~\ref{fig:suppression}d). For example, in superlattices and heterostructures of BiFeO$_3$ with centrosymmetric $Pnma$ LaFeO$_3$, BiFeO$_3$ has been reported to adopt the antiferroelectric PbZrO$_3$ structure~\cite{Carcan_et_al:2017,Carcan_et_al:2018}, or even observed in an entirely new antiferroelectric structure, which has not been reported in the bulk \cite{Mundy_et_al:2020}. 
Density functional calculations indicated that, for the strain conditions of the sample, this antiferroelectric phase is only slightly higher in energy than the ground-state polar phase, and it is favored because of its lower electrostatic energy cost \cite{Mundy_et_al:2020}. (Note that a DFT calculation for a BiFeO$_3$/LaFeO$_3$ slab in vacuum suggested the formation of a metallic layer at the interface, although that study did not explore the formation of non-polar BiFeO$_3$ phases and it is unclear how the polar discontinuties at the surface were treated in the calculation~\cite{sun_jpcc_2019}.)

\begin{figure}[ht]
\centering
 \includegraphics[scale=0.35, trim={1cm 6.5cm 3.0cm 0cm}, clip]{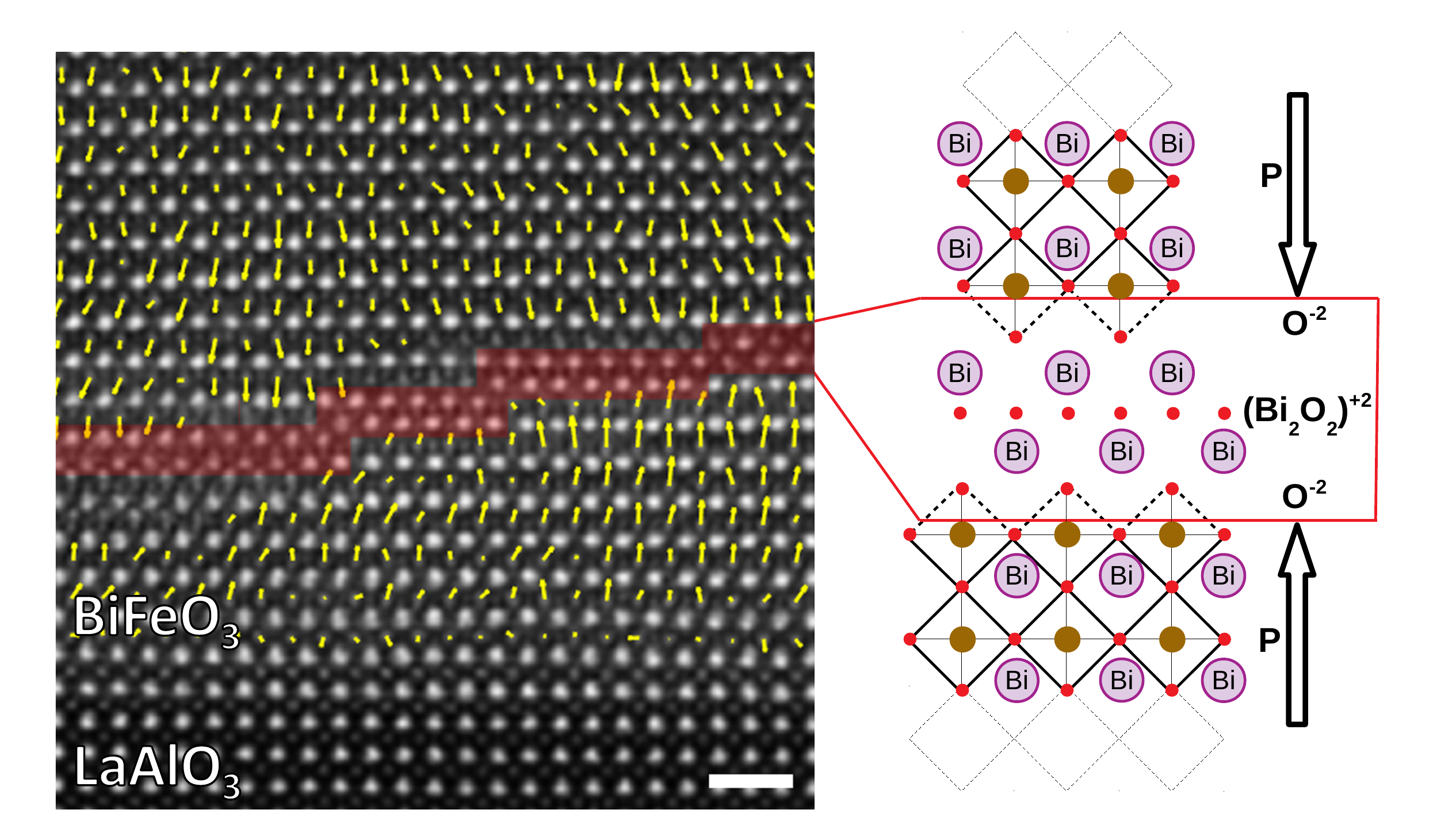}
 \caption{Cross-sectional HAADF-STEM image showing the presence of planar defects just above the BiFeO$_3$/LaAlO$_3$ interface. The stepped Bi$_2$O$_{2}$ structural units are highlighted by the red rectangles and the overlaid vector map of the ferroelectric polarization plotted opposite to the displacement of the Fe cations (yellow arrows) reveals that the polarization in the pseudo-tetragonal BiFeO$_3$ lattice changes from up to down direction across the defect. An atomic model of the defect is shown on the right. The scale bar is 1 nm.}
 \label{fig:lao_bfo}
\end{figure}

Another route to the compensation of the polar discontinuity is the creation of extended defects; we discuss the example of LaAlO$_3$/BiFeO$_3$, where this behavior has been observed, next. 
Under strong biaxial compressive strain, imposed by a small-lattice-constant substrate such as LaAlO$_3$, BiFeO$_3$ is known to undergo a phase transition to a tetragonal or tetragonal-like phase (T-BiFeO$_3$) with a large $c/a$ ratio of $\sim 1.3$ and a giant, almost entirely out-of-plane spontaneous polarization of  $\sim 150$ $\mu$C cm$^{-2}$ \cite{Zeches_et_al:2009,Hatt/Spaldin/Ederer:2010,Rossell-PRL:2009}. This spontaneous polarization is roughly three times the [001] spontaneous polarization of the usual rhombohedral phase of BiFeO$_3$, and correspondingly roughly three times the half-polarization quantum. (Note that the half-polarization quantum for T-BiFeO$_3$ is slightly larger, at 59 $\mu$C/cm$^2$, than that of the usual rhombohedral phase, because of its different lattice parameters. We obtain values of $a=b=3.67$ {\AA} and $c=4.64$ {\AA} in our calculations for the lowest energy tetragonal structure.) This giant spontaneous polarization has two implications: First, the spontaneous polarization can be at best only partially compensated by the layer polarization at a flat and stoichiometric BiO or FeO$_2$ (001) surface. 
Second, the giant polarization means that the electrostatic potential diverges strongly at an interface, and only a few layers can form before a compensation mechanism is required. 
In Fig.~\ref{fig:lao_bfo} we show a HAADF-STEM image of a 100nm-thick film of T-BiFeO$_3$ on LaAlO$_3$, in which, we observe such a compensation mechanism in the formation of an extended planar defect just a few unit cells above the T-BiFeO$_3$/LaAlO$_3$ interface. We indicate the local ferroelectric polarization (plotted opposite to the atomic displacements of the Fe cations) by the yellow arrows in  Fig.~\ref{fig:lao_bfo}; this vector map reveals that the first five to seven T-BiFeO$_3$ unit cells above the T-BiFeO$_3$/LaAlO$_3$ interface develop a spontaneous up-pointing polarization. 
Then, perhaps unexpectedly, above the extended planar defect the polarization in the T-BiFeO$_3$ lattice is reversed and a down-pointing polarization forms. This results in a head-to-head polarization configuration, with a giant discontinuity of the {\it spontaneous} polarization of $\sim 300$ $\mu$C cm$^{-2}$. For both the top and bottom layers, however, the stoichiometric BiFeO$_3$ terminates with an FeO$_2$ layer, so the absolute polarization of each layer is reduced from the spontaneous polarization by half a quantum, to $\sim 100$ $\mu$C cm$^{-2}$. Note that this is the ``happiest'' configuration possible for T-BiFeO$_3$, which, we emphasize again, does not have the accidental cancellation between its spontaneous polarization and the half-polarization quantum seen in the rhombohedral ground state. Correspondingly, the {\it absolute} polar discontinuity between the two T-BiFeO$_3$ layers is reduced to the (still very large!) value of $\sim 200$ $\mu$C cm$^{-2}$.

The planar defect (highlighted with red rectangles in Fig.~\ref{fig:lao_bfo}) consists of a characteristic Bi$_2$O$_{2}^{2+}$ layer, which shifts the perovskite layers above the defect half a perovskite block along the orthogonal horizontal directions in the manner of an Aurivillius phase, surrounded by two O$^{2-}$ layers (see cartoon on the right side of Fig.~\ref{fig:lao_bfo}; the horizontal red lines indicate the boundary of the defect). The total stoichiometry of the defect consists of one formally Bi$_2$O$_{2}^{2+}$ block plus two O$^{2-}$ ions per surface BiFeO$_3$ unit cell, leading to a net defect charge of two electrons, or $\sim -100 \mu$C cm$^{-2}$, per primitive unit cell cross-sectional area. This is exactly half of the polar discontinuity, and so, as sketched in the lower panel of Fig.~\ref{fig:lao}, is exactly the layer charge needed for compensation. Note that oxygen non-stoichiometry in the Aurivillius-like layer, yielding Bi$_2$O$_{2 \pm \delta}$ rather than precisely Bi$_2$O$_{2}$, is likely, and will change the exact amount of compensating charge that it provides. Indeed, a similar planar defect, with a {\it double} Bi$_2$O$_{2}^{2+}$ layer, was observed in a BiFeO$_3$ film grown on an LaAlO$_3$ substrate in spite of an intermediate metallic electrode between the film and the substrate \cite{Li-NanoLett:2017}. In this case the BiFeO$_3$ layer adjacent to the electrode was non-polar, and that above the defect developed a downward-pointing polarization as in our example. 

Interestingly, a similar extended defect has been reported as a surface ``skin'' in BiFeO$_3$, in all cases when the polarization points in the upwards (towards the surface) direction ~\cite{kim_mm_2013,jin_sr_2017, Xie-AdvMater:2017}. In Ref.~\onlinecite{jin_sr_2017} a film of rhombohedral BiFeO$_3$ was grown in [001] orientation on an insulating DyScO$_3$ substrate, and two opposite domains, separated by a 180$^{\circ}$ domain wall were imaged using HAADF-STEM. The down-polarization domain had a pristine BiO surface and so was in the `happy' configuration. The up-polarization domain, which would have been in the `unhappy' configuration with an excess positive charge in its pristine form, had a capping layer of the negatively-charged Bi$_2$O$_2$ Aurivillius-type extended defect to compensate. This finding has clearly unfavourable implications for the switching of BiFeO$_3$ domains. In Ref.~\onlinecite{Xie-AdvMater:2017}, the BiFeO$_3$ film on TbScO$_3$ had domains of strongly suppressed polarization alternating with domains of enhanced upward-pointing polarization; the latter had the surface skin overlayer. Finally, we mention that similar Aurivillius structures have also previously been observed as intergrowths in rhombohedral BiFeO$_3$ thin films~\cite{Deniz-JMaterSci:2014}. It would be interesting to analyze the interfaces between the intergrowths and the surrounding BiFeO$_3$ regions to determine the nature of the interface chemistry and the polarization orientation in the context of the compensation principles discussed here. 

\section{Summary and Outlook}

In summary, we have reviewed how the spontaneous polarization associated with ferroelectricity combines with the layer polarization associated with the ionic charges of the lattice to determine the electrostatic stability of the surfaces and interfaces of insulators. We reminded the reader that the two contributions are conveniently treated on the same footing by the modern theory of polarization, allowing straightforward determination of the amount of bound charge at a general surface or interface. The bound charge is important for the design of thin-film heterostructures and for the stability of surfaces, since any non-zero surface or interfacial charge must be compensated to avoid divergence of the electrostatic potential.

After briefly discussing examples of materials with a spontaneous polarization but no layer polarization ([001]-oriented PbTiO$_3$ and BaTiO$_3$) and a material with no spontaneous polarization but a non-zero layer polarization ([001]-oriented LaAlO$_3$) we focused on the case of multiferroic BiFeO$_3$, which combines both spontaneous and layer contributions. BiFeO$_3$ is of particular interest, because the spontaneous and layer contributions to the polarization in the ground-state $R3c$ structure have the same size along the usual [001] growth direction, leading to combinations of polarization and surface termination that are uncharged and therefore electrostatically stable. The opposite combinations have double the surface charge of a ferroelectric with the same spontaneous polarization but with uncharged layers. These have been referred to as the happy and unhappy combinations in earlier work \cite{efe_jcp_2020}.

We considered three scenarios:   BiFeO$_3$ on a metallic substrate,  BiFeO$_3$ in a superlattice or heterostructure with a zero-layer-charge insulator, and finally a BiFeO$_3$/insulator superlattice or heterostructure in which the insulator has the same layer charges as the BiFeO$_3$. In each case we illustrated the different possible behaviors with examples from the literature, as well as with density functional calculations and HAADF-STEM analyses performed for this work. 

We can summarize the differences in behavior in the three cases as follows: i) Both the happy and the unhappy polarization orientations can be stabilized by a metallic electrode, although the unhappy case is higher in energy and has a lower spontaneous polarization. This is consistent with the known large electric-field exchange bias of BiFeO$_3$ films, and implies that symmetric switching of a BiFeO$_3$ capacitor will be difficult to achieve. ii) The interface of BiFeO$_3$ with a zero-charge-layer insulator behaves like a free BiFeO$_3$ surface, with the happy combination of polarization and surface termination strongly favored. The unhappy combination can be stabilized by strong band bending to generate metallic layers at the interface and/or polarization of the adjacent insulator. iii) The interface of BiFeO$_3$ with another III-III perovskite has a polar discontinuity equal to the spontaneous polarization, and so is equally energetically unfavorable for both orientations of the polarization. As a result of the rich Bi-Fe-O low energy phase space, many responses are possible, including stabilization of phases with zero out-of-plane polarization, and the formation of extended planar defects.

While the electrostatic concepts discussed in this paper are not new, we hope that their collection in this article will be helpful in guiding the design of BiFeO$_3$ and related thin films or heterostructures with targeted electrical properties, as well as in interpreting experimental observations.

\section{Methods}
\label{sec:Methods}

\textit{Density functional Theory.} Density functional calculations were performed within the periodic supercell approach using the VASP code~\cite{vasp_1, vasp_2, vasp_3, vasp_4}. 
We chose the PBEsol functional~\cite{pbesol} for all calculations because i) it gives a good band alignment between metallic SrRuO$_3$ and BiFeO$_3$ and no pathological situation arises~\cite{Stengel_et_al:2011}, and ii) it yields a paraelectric ground state for SrTiO$_3$.
In order to obtain a band gap for BiFeO$_3$ close to the experimental value the Hubbard $U$, in the Dudarev~\cite{Dudarev_et_al:1998} approach, was used with $U-J=4$ eV on the Fe $3d$ states, and $U-J=2.0$ eV on the Ti and Ru $d$ states.
Core electrons were replaced by projector augmented wave (PAW) potentials~\cite{paw}, while the valence states (5e$^-$ for Bi, 8e$^-$ for Fe, 6e$^-$ for O, 10e$^-$ for Sr, 4e$^-$ for Ti, 8e$^-$ for Ru) were expanded in plane waves with a cut-off energy of 500 eV. 
In all calculations the in-plane lattice parameters were set to that of SrTiO$_3$, $a_{\mathrm{PBE}_{\mathrm{sol}}}=3.90$ \AA , as it is the substrate commonly used in epitaxial growth of BiFeO$_3$ thin films.
The in-plane surface area is $\sqrt{2} a \times \sqrt{2} a$.
A Monkhorst-Pack $k$-point grid of ($5 \times 5 \times 1$) was used for all ionic relaxations, which had an optimization threshold on the forces of 0.01 eV/\AA .
For the density of states calculations,  Monkhorst-Pack $k$-point grid of 11 $\times$ 11 $\times$ 1 was used.
An antiferromagnetic G-type ordering was imposed in BiFeO$_3$, which gave a magnetic moment of 4.15 $\mu_{\mathrm{B}}$ per Fe ion in the bulk.
SrRuO$_3$ instead is ferromagnetic and the magnetic moment is 1.4 $\mu_{\mathrm{B}}$ per Ru ion in the bulk.

The unit cell-by-unit cell polarization along the [001] direction shown in Fig.~\ref{fig:BFOSROsummary} and~\ref{fig:sto_bfo} was calculated by computing the displacement of each ion from the high symmetry position and multiplying it by the Born effective charges from Ref.~\cite{Neaton_et_al:2005}.

\textit{Thin Film Growth.} The BiFeO$_3$/La$_{0.7}$Sr$_{0.3}$MnO$_3$ (BFO/LSMO) heterostructures shown in Fig.~\ref{fig:BiFeO$_3$onLSMO} were grown by pulsed laser deposition on SrTiO$_3$ (001) (STO) single crystal substrates. Before the growth, a buffered HF acid-etch and thermal treatment process was used to obtain fully TiO$_2$-terminated surfaces. The sample with the MnO$_2$-terminated (La$_{0.7}$Sr$_{0.3}$O-MnO$_2$-BiO-FeO$_2$) interface was designed by growing whole LSMO unit cells directly on the STO substrate, followed by the growth of the BFO layer. For the sample with the La$_{0.7}$Sr$_{0.3}$O-terminated (MnO$_2$-La$_{0.7}$Sr$_{0.3}$O-FeO$_2$-BiO) interface, 1.5 unit cells of SrRuO$_3$ (SRO) were deposited on STO to switch the termination of the STO from TiO$_2$ to SrO. The SRO layer was grown at 650 $^{\circ}$C in 100 mTorr of oxygen pressure. Both the LSMO and BFO layers were grown at 690 $^{\circ}$C in 150 mTorr of oxygen pressure. A postannealing process was carried out at 400 $^{\circ}$C under an oxygen ambient for 1 h to ensure the samples were fully oxidized. For additional information, see Ref.~\cite{Yu_pnas_2012}. 
The BFO thin film shown in Fig.~\ref{fig:lao_bfo} was grown by molecular beam epitaxy (MBE) on single-crystal substrates of (001) LaAlO$_3$ (LAO). The studied 100-nm-thick film showed the coexistence of two interspersed BFO phases: a rhombohedral-like (R) phase and a tetragonal-like (T) phase. Further growth and characterization details can be found in Ref.~\cite{Zeches_et_al:2009}.

\textit{Transmission Electron Microscopy.} Cross-sectional specimens for transmission electron microscopy analysis were prepared by mechanical polishing using a tripod polisher followed by argon ion milling until electron transparency. High-angle annular dark-field scanning transmission electron microscopy (HAADF-STEM) was carried out using the TEAM 0.5 microscope located at the National Center for Electron Microscopy (NCEM). The TEAM 0.5 is a FEI Titan 80-300 microscope equipped with a high-brightness Schottky-field emission X-FEG electron source, a source monochromator, a CEOS DCOR spherical-aberration probe corrector, and a CEOS CETCOR spherical-aberration image corrector. The microscope was operated at 300 kV, the probe semi-convergence angle set to 16.5 mrad (which yields a calculated probe size of 0.63 Å), and the annular semi-detection range of the HAADF detector calibrated at 45–290 mrad. This setting was chosen to allow for a sufficiently large depth of field in order to enhance the contrast of the atomic columns. The positions of the atomic columns were first fitted by means of a center of mass peak-finding algorithm, and subsequently refined by solving a least-squares minimization problem (using the Levenberg–Marquardt algorithm). This iterative refinement was carried out using a custom-developed script that makes use of 7-parameter two-dimensional Gaussians and allows  estimation of the atomic column peak positions with picometer precision~\cite{Yankovich-NatCommun:2014, campanini_nanolett_2018}. Then, polarization maps were calculated from the relative displacements of the two cation sublattices present in the ferroelectric perovskite-type structures with general formula ABO$_3$. Thus, the local ferroelectric polarization was calculated by measuring the polar displacement in the image plane of the B position from the center of mass of its four nearest A neighbors. Here, in the polarization maps derived from HAADF-STEM images, the polarization vectors are plotted opposite to the displacement of the B cations.


\begin{acknowledgments}

N. A. S. acknowledges funding from the European Research Council (ERC) under the European Union’s Horizon 2020 research and innovation programme grant agreement No 810451. I. E. and C. G. acknowledge the use of the Euler cluster managed by the HPC team at ETH Zurich. M. D. R. acknowledges support by the Swiss National Science Foundation under Project No
200021-175926, and is thankful to R. J. Zeches, P. Yu and R. Ramesh for the samples used in this study.

\end{acknowledgments}

\section*{Data Availability}

The data that support the findings of this study are available within the article.

\newpage
\bibliography{Nicola.bib,Chiara.bib,Marta.bib}

\end{document}